\documentclass{article}
\usepackage{hiph-art}
\usepackage{graphicx}
\usepackage{latexsym}
\usepackage{rotating}

\newcommand{\beq}{\begin{equation}}
\newcommand{\eeq}{\end{equation}}
\newcommand{\bqa}{\begin{eqnarray}}
\newcommand{\eqa}{\end{eqnarray}}

\def\rightmark{Weibel Instability in the Melting Color Glass Condensate }
\def\leftmark{P. Romatschke and R. Venugopalan}
 
\title{Signals of a Weibel Instability in the Melting Color Glass Condensate} 
\authors{ 
{Paul Romatschke$^1$ and Raju Venugopalan$^{2}$ %
\index{One, A.} 
\index{Two, A.} 
}\\[2.812mm]
{\normalsize
\hspace*{-8pt}$^1$ 
Fakult\"at f\"ur Physik,
Universit\"at Bielefeld,
D-33615 Bielefeld, Germany\\[0.2ex] 
\hspace*{-8pt}$^2$ 
Physics Department,
Brookhaven National Laboratory,
Upton, NY 11973, U.S.A.
}}

\abstract{
Based on hep-ph/0510121,
we discuss further the numerical study 
of classical SU(2) 3+1-D Yang-Mills equations for 
matter produced in a high energy heavy ion collision. 
The growth of the amplitude of fluctuations as $\exp{(\Gamma \sqrt{g^2\mu \tau})}$ (where $g^2\mu$ is a scale arising
from the saturation of gluons in the nuclear wavefunction) is shown to be robust over a wide range of initial amplitudes that 
violate boost invariance. We argue that this growth is due to a non-Abelian Weibel instability, the scale of which is set by  
a dynamically generated plasmon mass. We discuss the relation of 
$\Gamma$ to the prediction from kinetic theory.}
\PACS{
}
\begin{document}
 
\maketitle

\vspace*{-0.5cm}
\section{Introduction}\label{intro}

One objective of the experiments done at ultra-relativistic heavy-ion
colliders such as RHIC and, in future, the LHC, is to understand 
the properties of very hot and dense partonic matter in QCD. This requires understanding how the coherent wavefunctions of 
the incoming nuclei decohere, possibly forming a thermal Quark Gluon Plasma (QGP). At high energies, the small x (or wee) 
partons determine the properties of nuclear wavefunctions. Their properties can be formulated in an effective field theory 
called the Color Glass Condensate (CGC)~\cite{CGC}. A semi-hard scale
$Q_s(x) >> \Lambda_{\rm QCD}$, the ``saturation" scale~\cite{GLR},
arises naturally in this approach and grows with energy; weak coupling
techniques are therefore feasible. Furthermore , the small x wavefunctions of the incoming nuclei can be treated as classical fields 
with large occupation numbers~\cite{MV}. This enables the description of nuclear collisions in terms of solutions of classical Yang-Mills equations 
with the fields representing the small x partons, and light cone currents describing the hard valence partons~\cite{KMW}. The latter can be modeled as 
\beq
J^{\mu}=\delta^{\mu +} \rho_{1}(x_\perp) \delta(x^-)+
\delta^{\mu -} \rho_{2}(x_\perp) \delta(x^+),
\label{current}
\eeq
where the color charge densities $\rho_{1,2}$ of the two nuclei are
independent sources of color charge ($x^\pm=(t\pm z)/2$). The $\delta$-function sources ensure that the fields are boost invariant, namely, independent 
of the space time rapidity $\eta$, defined as 
$\eta = {\rm atanh{(z/t)}}$.
The Yang-Mills equations can therefore be expressed in terms of 
the two transverse directions (${\bf x}_\perp$) and the proper time $\tau$, defined as $\tau = \sqrt{t^2-z^2}$. The initial conditions can be determined by 
matching the Yang-Mills equation in the four light cone regions, at $\tau=0$, to determine self-consistently the fields in the forward light cone in 
terms of those before the collision. These latter classical fields can be computed analytically in the CGC framework. 

In the McLerran-Venugopalan model (MV)~\cite{MV} for large nuclei, the sources of color charge are Gaussian distributed
\beq
\langle \rho_{i}^a({\bf x}_\perp)\rho_{j}^b({\bf y}_\perp)\rangle =
g^4 \mu^2 \delta_{ij} \delta^{ab} \delta^{2}({\bf x_\perp}-{\bf y_\perp})\,,
\label{eq:sources}
\eeq
where $g^2\mu$ is the dimensionful momentum scale in the problem. This scale is closely related to the saturation scale $Q_s$ which, in the 
classical effective theory, is defined as $Q_s^2 = g^4\mu^2 N_c \ln(g^2\mu/\Lambda_{\rm QCD})/2\pi$. 
For the initial conditions corresponding to configurations of color
sources of  each of the two nuclei, the Yang-Mills equations can be solved numerically, 
and the final gauge field configurations averaged over the sources, to determine energy and number distributions~\cite{KV1}.  

It is clear that for all finite $\sqrt{s}$ the ansatz of
$\delta$-function sources in Eq.~(\ref{current}) 
has to be modified in order to implement the fact that the nuclei will
{\em not} be contracted into infinitely thin sheets. 
More important, however, are the effects of quantum corrections,
which may be of order unity over rapidity scales $\sim 1/\alpha_s$. These are not included in the MV model but arise from 
small x quantum evolution of the classical fields~\cite{CGC}. Consequently, one must deal with functions having a 
finite width in $x^\pm$, respectively. For a single nucleus it is still possible to solve the Yang-Mills
equations classically and obtain the Weizs\"acker-Williams 
fields. For two nuclei, however, the problem becomes more involved,
simply because the nuclei will interact for a finite time and
the single nucleus solutions before the collision will be distorted
during this time span. 
Ignoring the details of this process, the main difference with respect
to the cases considered so far \cite{KV1} will be the emergence of
rapidity fluctuations and consequently a 
breaking of the boost-invariance of the small x fields.
In what follows, we will concentrate on studying the effect of
rapidity fluctuations by
numerically solving the Yang-Mills equations after the collision
(based on Ref.\cite{PaulRaju1});
to keep
the analysis as simple as possible, we assume the initial distortions of
exact boost-invariance to be very small. The reasons for this are two-fold. One is to connect our 
results  to published results \cite{KV1}. The other reason, as we will discuss, is to study the effects of 
the Weibel instabilities over several decades in the magnitudes of amplitudes.

This work is organized as follows: In section \ref{Motivation} we discuss 
why this study is connected to the phenomenon of 
non-Abelian plasma instabilities (\cite{Stan,us,ALM,AMYlet,Dumitru,3dVlasov}),
before providing details
of our setup in section \ref{Setup}. Our results are presented in
section
\ref{results}.
\vspace*{-0.3cm}

\section{Motivation}
\label{Motivation}

At earliest times $\tau Q_s \le 1$, typical gluon occupation numbers
are large and thus the system is described most appropriately in terms
of nonlinear gluonic fields, which should be accessible by simulating
classical Yang-Mills dynamics \cite{KV1}. However, because of the
rapid longitudinal expansion, the gluon occupation number drops until
the non-linearities become so weak that the hard modes ($p\sim Q_s$)
can be described as on-shell particles. In this regime (which should be
reached for $\tau Q_s \ge 1$), the dynamics of the system is in terms 
of hard particles coupled to soft ($k\ll g Q_s$) fields, so a
Vlasov-type kinetic approach should be appropriate to describe the
system. Consequently, at times $\tau Q_s \sim 1$, one would expect 
both classical and kinetic theory descriptions to offer a fair
approximation of the system dynamics.

Another consequence of the longitudinal expansion is that the typical
longitudinal gluon momentum, for a fixed slice in rapidity, becomes smaller as $p_z \sim
1/\tau$. Since the transverse gluon momentum stays approximately
constant $p_\perp \sim Q_s$, the gluon distribution
function $f({\bf p})$ tends to become more and more
anisotropic (until scatterings become important at very late
times). 

Using a Vlasov approach it has been shown  
\cite{Stan,us,ALM,AMYlet,Dumitru,3dVlasov} that when expansion effects
are negligible, systems with an anisotropic momentum-space
distribution function are subject to the presence of so-called 
plasma-instabilities, with a typical exponential growth rate $\gamma$ 
proportional to the soft scale $m_\infty$,
\beq
\gamma\sim \frac{m_\infty}{\sqrt{2}}, \qquad m_\infty^2=g^2 N_c \int \frac{d^3
p}{(2 \pi)^3} \frac{f({\bf p})}{|{\bf p}|}
\eeq 
in the limit of very strong anisotropies \cite{ALM}.
These instabilities manifest themselves as
exponentially growing magnetic fields which in turn reduce the
momentum-anisotropy, both by transferring energy from hard to soft
excitations as well as by bending hard particle trajectories.

Because of the relation between classical field dynamics and
kinetic theory at times $\tau Q_s\sim 1$ conjectured above, 
one would expect to see some
manifestation of these instabilities when simulating classical
Yang-Mills dynamics. To simulate an expanding metric, we solve 
the Yang-Mills equations in $(\tau,\eta,{\bf x}_\perp)$ co-ordinates. In momentum space, the 
conjugate momenta are 
$(k^\tau,k_\eta,{\bf k}_\perp)$, respectively. The Yang-Mills 
fields for ``soft" $k_\eta$ modes will thus be sensitive to anisotropic
 distributions of modes in (${\bf k}_\perp$,$k_\eta$), 
thereby triggering an instability of the Weibel type. 
It was predicted by Arnold, Lenaghan and Moore~\cite{ALM} that 
in an expanding system
such an instability would grow as 
$\exp\left(\sqrt{\tau}\right)$ rather than $\exp(\tau)$.
As shown in 
Ref.~\cite{PaulRaju1}, this is precisely what happens. 
Below, we discuss in some detail the setup of the numerical problem
and some of the results.

\vspace*{-0.3cm}
\section{Setup}
\label{Setup}
In $A^{\tau}=0$ gauge, the gluonic part of the QCD action has the
form \cite{KV1}
\beq
S=\int d\tau d\eta dx_\perp \tau {\rm Tr}%
\left[\frac{F_{\tau \eta}^2}{\tau^2}+F_{\tau i}^2-\frac{F_{\eta
i}^2}{\tau^2}-\frac{F_{i j}^2}{2}+\frac{j_\eta
A_\eta}{\tau^2}\right]=\int d\tau d\eta dx_\perp {\mathcal L},
\label{action}
\eeq
where 
$F_{\mu \nu}=\partial_{\mu} A_{\nu}-\partial_{\nu} A_{\mu}+i g 
[A_{\mu},A_{\nu}]$ is the field strength in the fundamental
representation with $F^{\mu \nu}=F^{\mu \nu}_{a} \tau_a$ and 
$[\tau_a,\tau_b]=i f_{a b c} \tau_c$, ${\rm Tr}\ \tau^a
\tau^b=\frac{\delta_{ab}}{2}$. In the following we shall ignore effects of
the current $j_\eta$. This is justified if we limit ourselves
to a small region around $\eta=0$.
With this restriction in mind we derive the conjugate momenta
from the Lagrangian Eq.(\ref{action}), 
\beq
E_i=\frac{\partial {\mathcal L}}{\partial (\partial_\tau A_i)}=\tau
\partial_\tau A_i,\qquad E_\eta=\frac{\partial {\mathcal L}}{\partial
(\partial_\tau A_\eta)}=\frac{1}{\tau}
\partial_\tau A_\eta
\eeq
with which we construct the Hamiltonian density
\beq
{\mathcal H}=E_i(\partial_\tau A_i) +E_\eta(\partial_\tau A_\eta)-
{\mathcal L}
={\rm Tr}\ \left[\frac{E_i^2}{\tau}+\frac{F_{\eta i}^2}{\tau}+\tau
E_\eta^2 + \tau F_{x y}^2\right].
\label{contHamil}
\eeq
Here transverse coordinates $x,y$ have been collectively labeled by
the Latin index $i$. Using finally
$
\partial {\mathcal H}/\partial E_\mu=\partial_\tau A_\mu,\ 
\partial {\mathcal H}/\partial A_\mu=-\partial_\tau E_\mu\,,
$
Hamilton's equations for the fields and their conjugate momenta are 
$$
\partial_\tau A_i=\frac{E_i}{\tau}, \quad \partial_\tau A_\eta =
\tau E_\eta, \quad
\partial_\tau E_i =\tau D_j F_{ji}+\tau^{-1} D_\eta F_{\eta
i}, \quad
\partial_\tau E_\eta =\tau^{-1}D_j F_{j\eta}.
$$

Since it will be used in the following, we also introduce two relevant
components of the stress-energy tensor in $(\tau,x,y,\eta)$ coordinates,
\bqa
T^{xx}+T^{yy}&=&2 \tau\ {\rm Tr}\left[F^2_{xy}+E_\eta^2\right]\\
\tau^2 T^{\eta \eta}&=&\tau^{-1}\ {\rm Tr}\left[F_{\eta i}^2+E_i^2\right]
-\tau\ {\rm Tr}\left[F_{xy}^2+E_\eta^2\right].
\label{Tmunu}
\eqa

\vspace*{-0.3cm}
\subsection{Initial conditions -- Boost-Invariant Case}

In the case of exact boost invariance, one obtains the initial
conditions by matching the equations of motions before the collision
(when there are only undisturbed single nucleus solutions) at the
point $x^\pm=0$ and along the
boundaries $x^+=0$, $x^->0$ and $x^-=0$, $x^+>0$. Omitting the details
worked out in \cite{CGC,KV1}, the result for the fields and momenta
at time $\tau=0$ is 
\bqa
&{\cal
A}_i(x_{\perp})=\alpha_{1,i}(x_\perp)+\alpha_{2,i}(x_\perp)\nonumber
\qquad 
{\cal A}_\eta(x_{\perp})=0\nonumber &\\
&{\cal E}_{i}(x_{\perp})=0\nonumber \qquad 
{\cal E}_\eta(x_{\perp})=i g [\alpha_{1,i}(x_\perp),\alpha_{2}^i(x_\perp)],&
\label{exact}\nonumber\\
&\alpha_{1,2}^i=\frac{i}{g} U_{1,2} \partial^i U_{1,2}^\dagger, \qquad  
U_{1,2}={\cal P}\exp{(-i \int_{-\infty}^{x^{\mp}} d x_{\pm}^\prime
\Lambda_{1,2}(x_{\perp},x_{\pm}^\prime))}&
\eqa
where ${\cal P}$ denoting path-ordering,
$\Delta \Lambda_{1,2}(x_\perp,x^\pm)=-
\rho_{1,2}(x_\perp) \delta(x^\pm)$ and $\rho_{1,2}$ are to be determined
from Eq.(\ref{eq:sources}).
%

\subsection{Initial conditions including rapidity fluctuations}

Ignoring the details of the initial rapidity profile we simply start
with the boost-invariant field configuration and disturb it by
adding small random rapidity variations.
Specifically, one has
$
A_i = {\cal A}_i, \ A_\eta=0, \
E_i = \delta E_i, \  E_\eta= {\cal E}_\eta+\delta E_\eta,
$
with $D_i \delta E_i + D_\eta E_\eta = 0$ at initial time
$\tau=\tau_{init}$. 
The rapidity dependent functions $\delta E_i$, $\delta E_\eta$ are
constructed as follows: for $\delta E_i$ we draw 
random configurations 
$\delta \bar{E}_i(x_\perp)$ in the transverse plane, 
$
< \delta \bar{E}_i(x_\perp) \delta \bar{E}_i(y_\perp)> = 
\delta^2(x_\perp-y_\perp)
$
and subsequently multiply by a random function $f(\eta)=\partial_\eta
F(\eta)$ with dimensionless amplitude $\Delta\ll 1$, 
\beq
<F(\eta)  F(\eta^\prime) > = \Delta^2 \delta(\eta-\eta^\prime)
\eeq
to get
$
\delta E_i(x_i,\eta)=f(\eta) \delta \bar{E}_i(x_i);
$
$\delta E_\eta$ is then constructed as
$\delta E_\eta = - F(\eta) D_i \delta \bar{E}_i(x_i)$.

Thus, by construction, one has added random rapidity fluctuations of
amplitude $\Delta$ to the system which obey Gauss's law. An 
advantage of this construction is of course that one can use periodic boundary
conditions in the $\eta$ direction, which for a lattice simulation is
somewhat more convenient. 

\vspace*{-0.3cm}
\subsection{Lattice Simulations}

To simulate the system we discretize space-time and use an adapted
leap-frog algorithm to evolve the system in time 
\cite{KV1} (details will be given elsewhere \cite{PaulRaju2}). 
The lattice parameters (all of which are dimensionless) are 
\begin{itemize}
\item $N_\perp$, $N_\eta$ the number of lattice sites in the
transverse/longitudinal direction
\item $g^2 \mu a_\perp$, $a_\eta$,
the lattice spacing in the transverse/longitudinal direction 
\item $\tau_{init}/a_\perp$, the time at which the 3-dimensional
simulations are started
\item $\delta \tau$, the time stepping size
\item $\Delta$, the initial size of the rapidity fluctuations
\end{itemize}

Of these, only the combinations 
$g^2 \mu a_\perp N_\perp \equiv g^2 \mu L$ and $a_\eta N_\eta \equiv
L_\eta$ (which correspond to the simulated system dimensions)
have physical meaning; the continuum limit is approached by keeping
these fixed while sending $\delta \tau\rightarrow 0,\ 
g^2 \mu a_\perp \rightarrow 0,\ a_\eta\rightarrow 0$.
For the 3-dimensional simulations, we still have to choose a value for
$\tau_{init}$, which should be such that for $\Delta=0$ we stay very
close to the result from the 2-dimensional simulations 
(for all of which $\tau_{init}=0$). Thus, we set
\beq
\tau_{init}=0.05\, a_\perp,
\eeq
but have checked that our results stay the same when choosing
$\tau_{init}/a_\perp=0.025,0.1$, respectively.

\section{Evolution of rapidity-fluctuations and a Weibel instability in expanding matter}
\label{results}

\begin{figure}
\begin{center}
\includegraphics[width=0.45\linewidth,angle=-90]{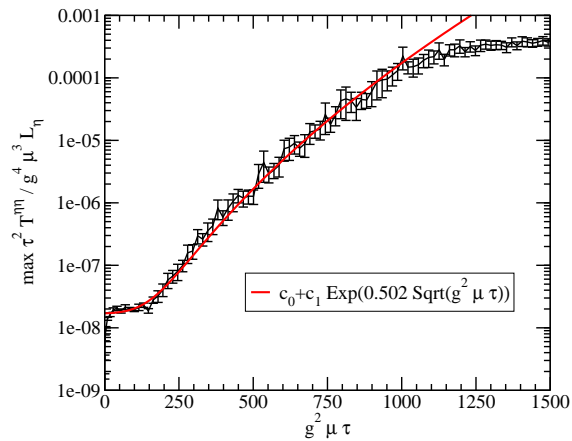}
\end{center}
\vspace*{-0.6cm}
\setlength{\unitlength}{1cm}
\caption{The maximum Fourier mode amplitudes of $\tau^2 T^{\eta \eta}$
for $g^2 \mu L=22.6$, $N_\perp=N_\eta=32$, $L_\eta=1.6$. Also shown
is a best fit with a $\exp{\sqrt{\tau}}$ behavior. 
The flattening out of the data at late times is due to the non-Abelian
interactions stopping the instability growth.
\vspace*{-0.3cm}}
\label{fig:fig1}
\end{figure}

An interesting quantity for the longitudinal dynamics is 
the energy momentum tensor component $T^{\eta \eta}$ (see Eq.(\ref{Tmunu})), 
of which we study the Fourier transform with respect to $\eta$,
\beq
\tilde{T}^{\eta \eta}(k_\eta,k_\perp=0)=\int d \eta\, \exp(i \eta\,
k_\eta) <T^{\eta \eta}(x_\perp,\eta)>_\perp,
\eeq
where $<>_\perp$ denotes averaging over the transverse coordinates $(x,y)$.
Apart from $k_\eta=0$, this quantity would be strictly zero in the
boost-invariant ($\Delta=0$) case. We have checked that this is indeed the case. For finite $\Delta$, however, 
$\tilde{T}^{\eta \eta}$ is in general non-vanishing for arbitrary
$k_\eta$ and possesses a maximum amplitude for one specific
$k_\eta$. Determining this maximum amplitude for each time step and
averaging over many random initial conditions, we obtain the curve shown in Fig.\ref{fig:fig1}. The curve is 
for $g^2\mu L=22.6$. In contrast to the curve shown in Ref.~\cite{PaulRaju1} (for $g^2\mu L = 67.9$), where 
the initial seed (violating boost invariance) was chosen to be very small ($\Delta \simeq 10^{-11}$), the seed here is 
4 orders of magnitude larger ($\Delta\simeq10^{-7}$). Nevertheless, for the same $g^2\mu L$ (see Table 1 in Ref.~\cite{PaulRaju1}-also 
included below), 
the fit to the growth rate $\Gamma$ is consistent with the result quoted in Ref.~\cite{PaulRaju1}. 

Another feature of our simulations that was not clear from our simulations with very small seeds was the flattening of the amplitude, the 
onset of which is seen in the Figure at $g^2\mu\tau \approx 1000$. 
We have checked (by varying lattice spacing by a factor of 8) that this 
phenomenon is rather insensitive to the ultraviolet modes, and 
appears to be a consequence of the ``non-Abelianization" of the
amplitude. In other 
words, the instability is cut-off when the non-Abelian self
interactions of the soft modes become important (also seen in 
Hard-Loop simulations without expansion \cite{3dVlasov}). 
More quantitative studies of this 
phenomenon will be presented in ~\cite{PaulRaju2}.

From Fig.\ref{fig:fig1}, one can see that from $g^2\mu \tau \approx 150$
onwards, there is rapid growth for which a best fit (up to times
$g^2 \mu \tau \sim 1000$) to the functional 
form $c_0 + c_1 \exp({\Gamma_{\rm fit} \tau^{c_3}})$ 
gives $\Gamma_{\rm fit} =0.502\pm 0.01$ for $c_3 = 0.5$; the
coefficients 
$c_0$, $c_1$ are small numbers proportional to the initial seed. 

In the presence of a Weibel
instability, one expects  $\tau^2 \tilde{T}^{\eta \eta}$ to grow 
as $\exp(\sim \gamma \tau)$.
For a system without expansion, $\gamma\sim m_\infty$ does not
change as a function of $\tau$ and thus 
the instability manifests itself through modes growing as 
$\sim\exp(\tau)$. However, as argued in Ref.\cite{ALM}, the soft scale $m_\infty$ behaves as
$m_\infty^2\sim 1/\tau$ in an expanding system. Therefore, the
functional form of the growth is changed to $\exp(\sqrt{\tau})$.
Our results confirm that this functional form is favored by a best fit
to our data in Fig.\ref{fig:fig1}.

We can confirm this interpretation by also
determining $m_\infty$ directly in our simulation. This is done by 
calculating\footnote{
The effect of small longitudinal fluctuations on transverse quantities
should be rather small.
Thus, we calculate this mass gap from a 2+1-dimensional simulation rather
than in the 3+1-dimensional case out of computational convenience.
}
the mass gap $\omega({\bf k_\perp}=0)$ of the gluon dispersion 
relation
defined as \cite{KV1}, 
\beq
\omega({\bf k}_\perp) = {1\over \tau}\sqrt{{{\rm Tr} \left[ E_i({\bf k}_\perp E_i(-{\bf k}_\perp) + \tau^2 E_\eta({\bf k_\perp}) E_\eta(-{\bf k}_\perp)\right]
\over {\rm Tr}\left[ A_i({\bf k}_\perp) A_i (-{\bf k}_\perp) +
{\tau}^{-2} A_\eta({\bf k}_\perp) A_\eta(-{\bf k}_\perp)\right]}},
\label{omdef}
\eeq
which should be proportional to the 
soft scale $\omega({\bf k_\perp}=0)\sim m_\infty$. We find \cite{PaulRaju1} 
$\omega({\bf k_\perp}=0)=\kappa_0\sqrt{g^2 \mu/\tau}$,
 consistent with the
expectation from \cite{ALM,BMSS}. Interpreting 
$\omega({\bf k_\perp}=0)$ as the plasmon
mass $\omega_{pl}$ we make use of the relation 
$\omega_{pl}^2=2/3 m_\infty^2$~ \cite{PaulRaju1} to obtain the quantitative
estimate $m_\infty=\kappa_0 \sqrt{3 g^2 \mu /(2 \tau)}$ 
for our simulation.

If we take the growth rate in the static case
and make the change $\gamma_{stat} \tau \rightarrow \gamma(\tau) \tau$
with $\gamma(\tau) = m_\infty(\tau)/\sqrt{2}$ for the expanding
system, we can define the ``theoretical'' growth
rate $\Gamma_{\rm theory}\sqrt{g^2\mu\,\tau}=2 \gamma \tau$.
Obtaining $\Gamma_{\rm fit}$ by a best fit to the data (e.g. in
Fig.\ref{fig:fig1}) for different values of $g^2 \mu L$, we can
compare this result to $\Gamma_{\rm theory}$, finding
\begin{table}[h]
\begin{center}
\begin{tabular}{|c|c|c|}
\hline
{\bf $g^2 \mu L$} & $\Gamma_{\rm theory}=\sqrt{3}\, \kappa_0$ &
$\Gamma_{\rm fit}$\\
\hline
22.6 & $0.526\pm0.003$ & $0.502 \pm 0.01$ \\
\hline
67.9 & $0.447\pm0.003$ & $0.427 \pm 0.01$\\
\hline
90.5 & $0.49\pm0.004$ & $0.46  \pm 0.04$\\
\hline
\end{tabular}
\end{center}
\end{table}

\vspace*{-0.5cm}
However, a
consistent treatment would require
that the growth rate in the expanding
case is $exp ( 2 \int_0^\tau  d\tau^\prime \gamma (\tau^\prime) )$; if
we assume as previously that $\gamma(\tau^\prime) =
m_\infty(\tau)/\sqrt{2}$, one obtains an additional factor of 2 in
the ratio of $\Gamma/\kappa_0$ relative to $\Gamma_{fit}$.
Understanding this difference of a factor of $2$ requires a more careful
study of the correspondence between the dynamics of expanding fields in our
case and
that in the static HTL case. In particular, it is important to investigate
how $\omega_{pl}$ that we extract from the lattice relates to the plasmon
frequency in Hard Thermal Loop simulations. Studies in this direction are
in progress.
\vspace*{-0.4cm}
\section*{Acknowledgments}
RV's research is supported by DOE Contract
No. DE-AC02-98CH10886. He thanks the Alexander von Humboldt
foundation for support during the early stages of this work.
PR was supported by DFG-Forschergruppe EN164/4-4. We would like to thank 
D.~B\"odeker, J.~Engels, F. Gelis, D. Kharzeev, A. Krasnitz, M.~Laine,
T. Lappi, L. McLerran, Y. Nara, R. Pisarski and M. Strickland 
for fruitful discussions. PR  thanks the organizers
of ``Quark-Gluon-Plasma Thermalization'' for this particularly 
nice workshop.

\vfill\eject
\end{document}